\documentclass{article}

\usepackage{microtype}
\usepackage{graphicx}
\usepackage{subfigure}
\usepackage{booktabs} 
\usepackage{amsmath}
\usepackage{amsfonts,amssymb}
\usepackage{algorithm}
\usepackage{algorithmic}
\usepackage{hyperref}
\usepackage{multirow}

\def\mU{{\mathcal U}}

\DeclareMathAlphabet\mathbfcal{OMS}{cmsy}{b}{n}

\def\0{{\bf 0}}
\usepackage{ dsfont }
\def\1{\mathds{1}}

\usepackage{ntheorem}

\usepackage{enumitem}

\usepackage[accepted]{icml2020}

\begin{document}

\icmltitlerunning{\fontsize{8.0pt}{\baselineskip}\selectfont{A Thorough Comparison Study on Adversarial Attacks and Defenses for  Common Thorax Disease Classification in Chest X-rays}}

\twocolumn[
\icmltitle{A Thorough Comparison Study on Adversarial Attacks and Defenses for  Common Thorax Disease Classification in Chest X-rays}

\icmlsetsymbol{equal}{*}

\begin{icmlauthorlist}
\icmlauthor{Chendi Rao}{scut}
\icmlauthor{Jiezhang Cao}{scut}
\icmlauthor{Runhao Zeng}{scut}
\icmlauthor{Qi Chen}{scut}
\icmlauthor{Huazhu Fu}{Abu}
\icmlauthor{Yanwu Xu}{Cixi}
\icmlauthor{Mingkui Tan}{scut}
\end{icmlauthorlist}

\icmlaffiliation{scut}{School of Software Engineering, South China University of Technology, GuangZhou, China}
\icmlaffiliation{Abu}{Inception Institute of Artificial Intelligence, Abu Dhabi, UAE}
\icmlaffiliation{Cixi}{Cixi Institute of Biomedical Engineering, Chinese Academy of Sciences, Ningbo, China}
\icmlcorrespondingauthor{Mingkui Tan}{mingkuitan@scut.edu.cn}

\vskip 0.3in
]
\printAffiliationsAndNotice{}

\begin{abstract}
Recently, deep neural networks (DNNs) have made great progress on automated diagnosis with chest X-rays images.
However, DNNs are vulnerable to adversarial examples, which may cause misdiagnoses to patients when applying the DNN based methods in disease detection.
Recently, there is few comprehensive studies exploring the influence of attack and defense methods on disease detection, especially for the multi-label classification problem.
In this paper, we aim to review various adversarial attack and defense methods on chest X-rays.
First, the motivations and the mathematical representations of attack and defense methods are introduced in details.
Second, we evaluate the influence of several state-of-the-art attack and defense methods for common thorax disease classification in chest X-rays.
We found that the attack and defense methods have poor performance with excessive iterations and large perturbations.
To address this, we propose a new defense method that is robust to different degrees of perturbations. 
This study could provide new insights into methodological development for the community.

\end{abstract}

\section{Introduction}
\label{Introduction}
Disease detection using X-rays can be categorized into radiologists diagnosis and DNNs based methods.
Radiologists diagnosis requests rich expert knowledge and experience.
Recently, DNNs based methods have achieved great process in x-rays analysis, including classification~\cite{wu2019deep}, segmentation\cite{cernazanu2013segmentation}, detection\cite{rajpurkar2017chexnet}.
Unfortunately, DNNs are vulnerable to adversarial examples \cite{kurakin2016adversarial2}, and thus make wrong predictions when introducing small and imperceptible perturbations to clean images.
Recently, many attack \cite{shi2019curls} and defense \cite{hu2019new} methods have been proposed on natural images for the multi-class classification problem.
However, analyzing attack and defense methods on medical images for multi-label classification task remains an open question.

Most attack methods focus on the multi-class classification task, but neglect to explore the performance on the multi-label classification task on medical images. 
Existing attack methods can be divided into three categories in medical deep learning systems, namely gradient-based methods, score-based methods, and decision-based methods. 
Here, the Gradient-based methods calculate the gradient of the specific algorithm with regard to the loss of adversarial examples iteratively. In addition,  gradient-based methods constrain the size of perturbations to ensure the perturbations are imperceptible to humans.
However, many fundamental questions have not been investigated.
For example, it is worth studying whether more iterations and larger perturbation affect the performance of the attack methods.

Some defense methods were proposed to defense against the adversarial examples and improve the robustness of the model. 
However, these methods neglect the quality of the adversarial examples in the adversarial training. 
In practice, these methods are sensitive to the degrees of perturbation.
For example, gradient-based methods with different iterations and constrains will attack models to different degrees. 
Therefore, it is worth detecting whether the defense methods are able to defend against different degrees of perturbation.

In this paper, we provide a
comprehensive survey on attack and defense methods for the automated chest X-ray analysis systems. 
Specifically, we evaluate the performance of the attack and defense methods on chest X-ray images, \textit{i.e.}, CheXpert dataset~\cite{irvin2019chexpert}.
Moreover, we conduct experiments to explore the influence of more iterations and large perturbation for these methods.
Last, the future open research problems for both attack and defense methods are pointed out.

The contributions of this paper are summarized as follows: 
\begin{itemize}
\item We investigate the impact of the iterations and perturbations on the attack methods on the CheXpert dataset with multi-label classifiers. 
We found that superabundant iterations and oversized perturbations will damage the performance of attack methods.
\item We evaluate the robustness of the defense methods.
Moreover, we found that these methods are not robust to the adversarial examples with large perturbation for the multi-label classification.
\item We propose a new defense method to improve the robustness of the classifier. 
Moreover, the experiments demonstrate that the proposed method outperforms existing defense methods. 
\end{itemize}


\section{Related Work}

\textbf{Deep learning on chest X-rays.}
DNNs have made great progress in image analysis, such as \cite{zhang2019whole,guo2019nat,zhang2019collaborative}.
The chest X-ray is an important tool to detect diseases such as cardiomegaly, pneumothoraxes, and consolidation. Many works have been done using deep neural networks to detect diseases from chest X-rays. \cite{lakhani2017deep} showed a method to automatically detect pulmonary tuberculosis using convolutional neural networks. \cite{pasa2019efficient} proposed an efficient method to detect tuberculosis based on a deep neural network with less computational and memory requirements than those of other methods. \cite{stephen2019efficient} designed a specific neural network architecture for pneumonia classification. \cite{saul2019early} proposed an architecture that consists of a convolutional neural network and residual network for the pneumonia classification task. \cite{wang2018chestnet} presented the ChestNet to automatically diagnose thorax diseases using chest radiography, including Atelectasis, Cardiomegaly, Effusion, and other four diseases.

\noindent\textbf{Attack methods.}
\cite{szegedy2013intriguing} showed that classifiers may make the wrong prediction with high confidence when suffering some imperceptible perturbations. They also found models are vulnerable to the adversarial examples generated by other models, which is called the transferability of adversarial examples. \cite{kurakin2016adversarial} demonstrated that in the physical world, classifiers are still vulnerable to adversarial examples, and they proposed a single-step method and an iterative method to generate adversarial examples. Based on these studies, many methods have been proposed to improve the performance of adversarial examples. \cite{brown2017adversarial} proposed a method that creates universal adversarial image patches in the real world. \cite{su2019one} showed a method that fools deep neural networks using one pixel, and \cite{shi2019curls} proposed a method that reduces the noise of adversarial examples and improves the transferability of adversarial examples; \cite{moosavi2016deepfool} introduced Deepfool, which computes a minimal norm adversarial perturbations for a given image in an iterative manner. 
\begin{table}[!t]
	\centering
    \resizebox{0.5\textwidth}{!}{
	\begin{tabular}{c|c|c}
		\hline
		Methods  &    Step (s)    &                  Characteristic                  \\ \hline\hline
		  FGSM   & single-step &                with no iterations                \\ \hline\hline
		  PGD    & multi-step & with a random start using a uniform distribution \\ \hline\hline
		 MIFGSM  & multi-step  &       add momentum into the attack process       \\ \hline\hline
		DII-FGSM & multi-step &      transform images with probability $p$       \\ \hline\hline
		  DAA    & multi-step &     apply kernel function to attack process      \\ \hline
	\end{tabular}
	}
 \caption{Comparisons of different attack methods.}
 \label{table:table 1}
\end{table}

\noindent\textbf{Defense methods.}
Adversarial examples threaten the security of DNNs. Therefore, many methods were proposed to defend against adversarial examples and improve the robustness of models. \cite{hinton2015distilling} introduced the concept of distillation to resist against adversarial examples; \cite{kurakin2016adversarial2} proposed to use adversarial training to defense by augmenting the training set with both original and perturbed data using a single-step attack method; \cite{tramer2017ensemble} further introduced ensemble adversarial training, a technique that augments training data with perturbations transferred from other models. \cite{guo2017countering} demonstrated to transform adversarial images using cropping, quilting or total variance minimization. \cite{samangouei2018defense} proposed to protect classifiers against adversarial attacks using Generative Adversarial Nets. \cite{li2019improving} conducted adversarial training using triplet loss to improve the robustness of models. \cite{hu2019new} regarded the omnipresence of adversarial perturbations as a strength rather than a weakness to achieve a high accuracy under the white-box setting. These methods influence DNNs in different aspects and show different performance against attack methods. 

\section{ Attack and Defense Methods}
The goal of this paper is to review various adversarial attack and defense methods on chest X-rays.
For attack methods, we introduce five popular gradient-based methods. For defense methods, we discuss a traditional adversarial training strategy and a pixel deflection transform training strategy. Moreover, we propose a novel defense method to improve the defense performance. 

\subsection{Attack Methods}

Given a classifier $f$ : $x$ $\in$  $\mathcal{X}$ $\rightarrow$ $y$ $\in $ $\mathcal{Y}$, the attack methods aim at adding perturbations to a clean image $x$ to obtain a new adversarial sample $x^*$, which is misclassified by the classifier and imperceptible to humans.
In this paper, we introduce five popular gradient-based attack methods as shown in Table \ref{table:table 1}.
These methods satisfy the $L_\infty$ norm bound to restrict the size of the perturbations and solve the constrained optimization problem as follows:
\begin{equation}
\mathop{\arg\max}\limits_{x^*}J(\theta,x^*,y),~~\mathrm{  s.t.}~ ||x^* - x||_\infty \leq \epsilon,
\end{equation}
where $J(\theta,x^*,y)$ is a loss function, and $\theta$ is the parameter of the model.
For clarity, we formulate the adversarial examples of the gradient-based methods as follows:
\begin{equation}
x_{t+1}^{*} = Clip_{x}^{\epsilon} \{x_{t}^{*} +  \alpha \cdot \text{sign}(G_{t+1}(\theta, x_{t}^{*}, y))\},
\end{equation}
where $t \geq 0$, $G_{t+1}(\theta, x_{t}^{*}, y )$ represents a certain function generated through the loss about clean images, and $x_{t}^{*} $ represents an image generated by attack algorithm in t step,   $Clip_{x}^{\epsilon}$ clips the pixel values  to ensure they are in the $\epsilon$-ball of the original image $x$. 
In the following, we will discuss the motivations and mathematical representations of five gradient-based methods.
 
\paragraph{Fast Gradient Sign Method (FGSM)} \cite{kurakin2016adversarial} is a one-step attack method.
It adds the sign of the gradient of the loss function $J$  with regard to the image $x$ to the original image. 
In FGSM, 
\begin{align}
G_{t+1}(\theta, x^{*}_{t}, y) = \nabla_x J(\theta,x,y),
\end{align}
where $t = 0$.
FGSM is a simple method that does not require any iterative procedure to compute the gradient. It is faster than other multi-step methods. In addition, it just calculates the gradient once, and so the success rates often lower than that of other methods.

\paragraph{Project Gradient Descent (PGD)}~\cite{madry2017towards} is a multi-step method based on FGSM with a random start through a  uniform distribution $\mU$ between -$\epsilon$ and   $+\epsilon$ for the original image. Therefore, we set $  x^{*}_{0} = x + \mU(-\epsilon,+\epsilon) $,  the $G_{t+1}(\theta,x_{t},y)$ can be described as follows:
\begin{align}
        G_{t+1}(\theta,x^*_{t},y)&=\nabla_x J(\theta,x^*_{t},y). 
\end{align}

Although PGD achieves higher success rates than FGSM, it is more time-consuming since it calculates the gradient for several iterations.
~~\cite{madry2017towards} shows that models robust to PGD are always robust against attacks that rely on first-order information.

\paragraph{Momentum Iterative Gradient-based Method (MIFGSM)} \cite{dong2018boosting} is an iterative variant of FGSM. MIFGSM applies the momentum into the attack process to improve the transferability of adversarial examples. In MIFGSM, we initialize $x_{0}^{*} = x$, the $G_{t+1}(\theta,x_{t}^{*},y)$ can be defined as follows:
\begin{equation}
G_{t+1}(\theta,x^*_{t},y){=}\left\{
\begin{array}{lcl}
\mu G_{t}(\theta,x^*_{t-1},y) {+} g_t(J), &{t \geq 1},\\
g_0(J), & {t=0},\\
\end{array} \right.
\end{equation}
where $g_t(J):=\frac{\nabla_x J(\theta,x^*_t,y)}{||\nabla_x J(\theta,x^*_t,y)||_{1}}, t\ge0$.
MIFGSM maintains high success rates for white box attacks and also improves the success rates for black box attacks.

\paragraph{Distributionally  Adversarial Attack (DAA)}~\cite{zheng2018distributionally} solves the problem of optimal adversarial examples. The DAA satisfies the $L_\infty$ constrain and it increases the maximum generalized risk by adding perturbations to clean images.In DAA, we set $x^{i}_{0} = x^i$, and the $G_{t+1}(\theta,x_{t}^{i},y)$ can be depicted as follows:
\begin{align}
&G_{t+1}(\theta,x^i_{t},y) =\nabla_{x^i_t}J(\theta,x^i_t,y^i)\\
&+\frac{c}{M}\left[\mathop{\sum}\limits_{j = 1}^M K(x^i_t,x^j_t)\nabla_{x^j_t}J(\theta,x^j_t,y^j)
+\nabla_{x^j_t}K(x^i_t,x^j_t)\right] \nonumber,
\end{align}
where $K(\cdot,\cdot)$ is a kernel function such as the RBF kernel, c is a hyperparameter, $M$ is the minibatch size, and $x^{i}_{t}$ represents adversarial examples. DAA is a strong attack method that outperforms the PGD in both white box attacks and black box attacks. However, it spends a lot of time getting the gradient of the kernel function.

\paragraph{Diverse Input Iterative Fast Gradient Sign Method(DII-FGSM)}~\cite{xie2018improving} randomly transforms
input images with probability $p$ and maximizes the loss function  with regard to these transformed inputs. The $G_{t+1}(\theta,x_{t}^{*},y)$ can be define as
\begin{equation}
G_{t+1}(\theta,x^*_{t},y) = \nabla_x J(\theta,\mathcal{T}(x^*_{t};p),y).
\end{equation}
Here, the  transformation function  $\mathcal{T}(x^*_t;p)$ is as follows:\\
\begin{equation}
\mathcal{T}(x^*_t;p)=\left\{
\begin{array}{lcl}
\mathcal{T}(x^*_t),      &      & \mathrm{with~probability} ~p,\\
x^*_t,   &      & \mathrm{with~probability} ~1-p.\\

\end{array} \right.
\end{equation}
Specifically, we set the transformation as image resizing.

DII-FGSM improves the transferability of adversarial examples.  It outperforms the MI-FGSM when attacking ensemble models in the black box attacks.


\subsection{Defense Methods}

In this paper, we introduce a defense method that trains models with clean images and adversarial examples and another method that changes the pixels of images to destroy the noise interference. In addition, we analyze the shortages of each method and propose a new defense method.


\paragraph{PGD adversarial training (Adv\_train)}~\cite{madry2017towards} is an attack method that fools models with high success rates.
\cite{madry2017towards} found that models will be robust against first-order attack methods when training the model with adversarial examples generated by PGD. 
In adversarial training, we train the models on the samples that combine clean images and adversarial examples, as recommended by~\cite{goodfellow2014explaining}
\begin{equation}
\mathop{\!\!\!\!\arg\min}\limits_{\theta}\left[\mathop{\mathbb{E}}\limits_{(x, y) \sim p_{xy}}\!\!\mathop{\max}J(\theta,x^*,y)+\!\!\mathop{\mathbb  E}\limits_{(x,y) \sim p_{xy}}\!\!J(\theta, x, y)\right],\!
\end{equation}
where $p_{xy}$ represents the distribution of training data.
The total loss can be defined as follows:
\begin{equation}
\mathcal{L}(\theta,x,y) = \lambda J(\theta,x,y) + (1-\lambda)J(\theta,x^*,y).
\end{equation}

PGD adversarial training generates adversarial examples in each epoch, it costs a lot of time to train model, and it retains the high accuracy of the clean images.

\paragraph{Pixel deflection transform (PDT)}~\cite{prakash2018deflecting} randomly samples a pixel from adversarial examples, and replaces the pixel with another pixel that was selected from a small square neighborhood. In addition, to soften the pixel deflection, PDT also applies some denoising operations.

PDT transforms images before inputting them into networks. It reduces the accuracy of the clean images, but defend against adversarial examples with different sized perturbations in both white box attacks and black box attacks. Furthermore, it does not require any training during this process, so it is more efficient than PGD adversarial training.
 
\paragraph{Our defense method.}
Adv$\_$train training models are able to keep a high accuracy of clean images, but do not robust to adversarial examples generated by large-sized perturbations (see Figure \ref{fig:fig6}). PDT changes the pixel of images to defense against adversarial examples. PDT is able to robust to large-sized perturbations but reduces the accuracy of clean images. In this sense, we combine the two methods together to improve performance.
Specifically, we first train a model using adversarial training, then transform images by PDT, and finally input the transformed image into the model to get the prediction. Our defense method gets higher accuracy of clean images than PDT and is robust to large-sized perturbations.  


\section{Experiments}

\subsection{Dataset and Evaluation Metric}
A chest X-ray is a kind of medical image that can be used to detect many lung ailments.
In the experiments, we test attack and defense methods on CheXpert ~\cite{irvin2019chexpert}, which is the largest chest X-ray dataset, consisting of  224,316 chest X-rays with 14 chest radiographic observations. To evaluate the methods, we select 6 observations of the dataset: No Finding,  Atelectasis, Cardiomegaly, Consolidation, Edema, and  Pleural Effusion.
The training labels in the dataset are 0 (negative), 1 (positive), or $u$ (uncertain). We classify the labels of Atelectasis and Edema as 1 and the other 4 observations as 0  when the label is $u$. 
 
We use the area under curve (AUC) as \textbf{evaluation metric}.  The value of AUC ranges from 0 to 1. A perfect model would approach to 1 while a poor model would approach to 0.
 

\subsection{Implementation Details}
\paragraph{Training details.}
In the experiments, we train networks using clean images and samples that combine clean images and adversarial examples. We train Densenet121~\cite{huang2017densely},
Resnet50~\cite{he2016deep}, VGG16~\cite{simonyan2014very}, and Inception v3~\cite{szegedy2016rethinking} using clean images and conduct adversarial training using Densenet121. During  the training process, we use the Adam optimizer with $\beta_1$ = 0.9 and $\beta_2$ = 0.999 and set the learning rate as 0.0001. In Table \ref{tabel:tabel2}, we show the details about settings of the networks.
 \begin{table}[!t]
	\centering

     \setlength{\tabcolsep}{2.5mm}
	\begin{tabular}{c|c|c|c}
        \hline
         Model &  Parameters &Image Size & AUC    \\
        \hline
        \hline
        Densenet121 & 7.98M&224 & 0.87806  \\
        \hline

         \hline
         Resnet50& 25.56M&224 &0.87370   \\
        \hline

        \hline
          VGG16&138.36M & 224 & 0.88378 \\
        \hline

        \hline
          Inception v3&24.7M & 299&0.82679   \\
        \hline
	\end{tabular}
\caption{Settings of different models, \textit{i.e.,} number of parameters and size of input. We also report the average AUC of each model.}
 \label{tabel:tabel2}
\end{table}

\paragraph{Settings of attack methods.}
To evaluate the performance of the attack methods, we conduct our experiments by attacking a single model and attacking ensemble models using four networks trained by clean images, including Densenet121 (Den-121), Resnet50 (Res-50), VGG16 and Inception v3 (Inc-v3) in black box attacks and white box attacks. 

In order to attack ensemble models, We generate adversarial examples using an ensemble of three models and evaluate the methods using the ensemble models and the hold-out model. There are three ensemble methods introduced by ~\cite{dong2018boosting}. In our experiments, we ensemble models together using ensemble in logits. It can be described as
\begin{equation}
     l(x) = \Sigma^{K}_{k = 1}\omega_kl_k(x),
\end{equation}
 where $l_k(x)$ are the logits of the k-th model, $\omega_k$ is the weight where $\omega_k\geq0$ and $\Sigma^{K}_{k=1}\omega_k = 1$. In our works, we assign each model the same weight, and thus  we set  $\omega_k = \frac {1}{3}$.
In practice, we set $\epsilon$ = 0.3 to restrict the size of the perturbations of each attack method. 
We input the logits into a sigmoid function and calculate the binary cross-entropy loss at last.


\paragraph{Settings of defense methods.}

\begin{figure}[t]
    \centering
    \includegraphics[scale=0.42]{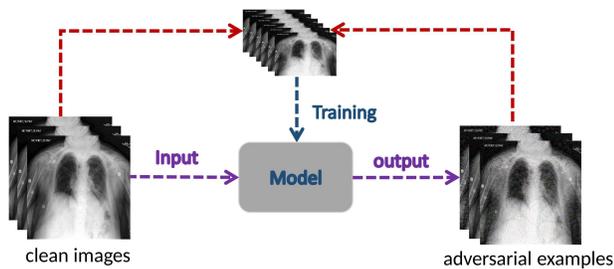}
    \caption{The process of adversarial training.  First, we input clean images into model to generate adversarial examples. Then we train model using clean images and adversarial examples.}
    \label{fig:fig 3}
\end{figure}

To evaluate the defense methods, we generate adversarial examples using PGD for adversarial training.
The adversarial training process is shown in Figure \ref{fig:fig 3}. In our experiments, we generate adversarial examples using PGD with 10 iterations, and we set $\epsilon$ as 4/255.
We set $\lambda$ = 0.6 and define the loss function $J(\theta,x,y) $ as a binary cross-entropy loss.
During the adversarial training process, to improve the AUC of clean images, we pre-train the model using clean images with 6 epochs.

For the pixel deflection transform,
we deflect pixels of adversarial examples without the class activation map and conduct non-local means denoising for images. To test the performance of PDT, we generate adversarial examples using attack methods with 10 iterations and we set $\epsilon$ as 4/255.


\subsection{Attack on Multi-label Classification Task}

\begin{table}[t]
    \scriptsize
	\centering
     \setlength{\tabcolsep}{7pt}
	\begin{tabular}{c|c|c|c|c|c}
        \hline
        Model &  Attack & Den-121 & Res-50 & VGG16 & Inc-v3 \\
        \hline
		\hline
		 & PGD&$\textbf{0.02704 }$ &0.28013  & 0.33482 &  0.45193 \\
         & MIFGSM &\textbf{0.02752 }  & 0.24133 & 0.31822 & 0.43765 \\
        Den-121 & DAA&\textbf{0.02665} & 0.25180 & 0.30802 & 0.41879 \\
         & DII-FGSM&\textbf{0.03194} & 0.19913&  0.27863& 0.43040 \\
         & FGSM&\textbf{0.29768}  & 0.48566 & 0.48811 & 0.49160 \\
		\hline

		\hline
		 & PGD& 0.16108&\textbf{0.04050}& 0.35482 & 0.40288  \\
         & MIFGSM & 0.14699&\textbf{0.03229}  & 0.32746 & 0.37085\\
        Res-50 & DAA&0.14512  &\textbf{0.02114} & 0.32061 & 0.39489 \\
         & DII-FGSM&0.12379&\textbf{0.04133} &0.28092& 0.36885\\
         & FGSM& 0.46772& \textbf{0.29976} &0.50658 & 0.47217 \\
		\hline

        \hline
		 & PGD&0.15223 &0.14941&\textbf{0.09472} &0.37467 \\
         & MIFGSM & 0.13291& 0.13353&\textbf{0.09281} &0.31479 \\
        VGG16 & DAA& 0.12333 & 0.13516 & \textbf{0.06760} & 0.34499 \\
         & DII-FGSM&0.10633 &0.10876 &\textbf{0.09185} & 0.35588 \\
         & FGSM& 0.48291 & 0.49232 &\textbf{0.33788} & 0.60106\\
		\hline

         \hline
		 & PGD& 0.54718 & 0.51072& 0.56739& \textbf{0.04509}  \\
         & MIFGSM &0.52933& 0.47201 &0.49618 &\textbf{0.04532} \\
        Inc-v3 & DAA&0.45780 &0.37316&0.38287 & \textbf{0.03989} \\
         &DII-FGSM&0.52348&0.47961 &0.49061& \textbf{0.05153} \\
         & FGSM& 0.47323 & 0.50655 & 0.52806&\textbf{0.20117}  \\
		\hline
	\end{tabular}
 \caption{Average AUC of untargeted attacks against different models. The adversarial examples are crafted for Den-121, Res-50, VGG16, and Inc-v3. \textbf{The bold blocks are white box attacks.}}
 \label{tabel:tabel3}
\end{table}

\begin{figure}[t]
\centering
\subfigure[]{
    \begin{minipage}[t]{0.5\linewidth}
        \centering
        \includegraphics[width=1\linewidth]{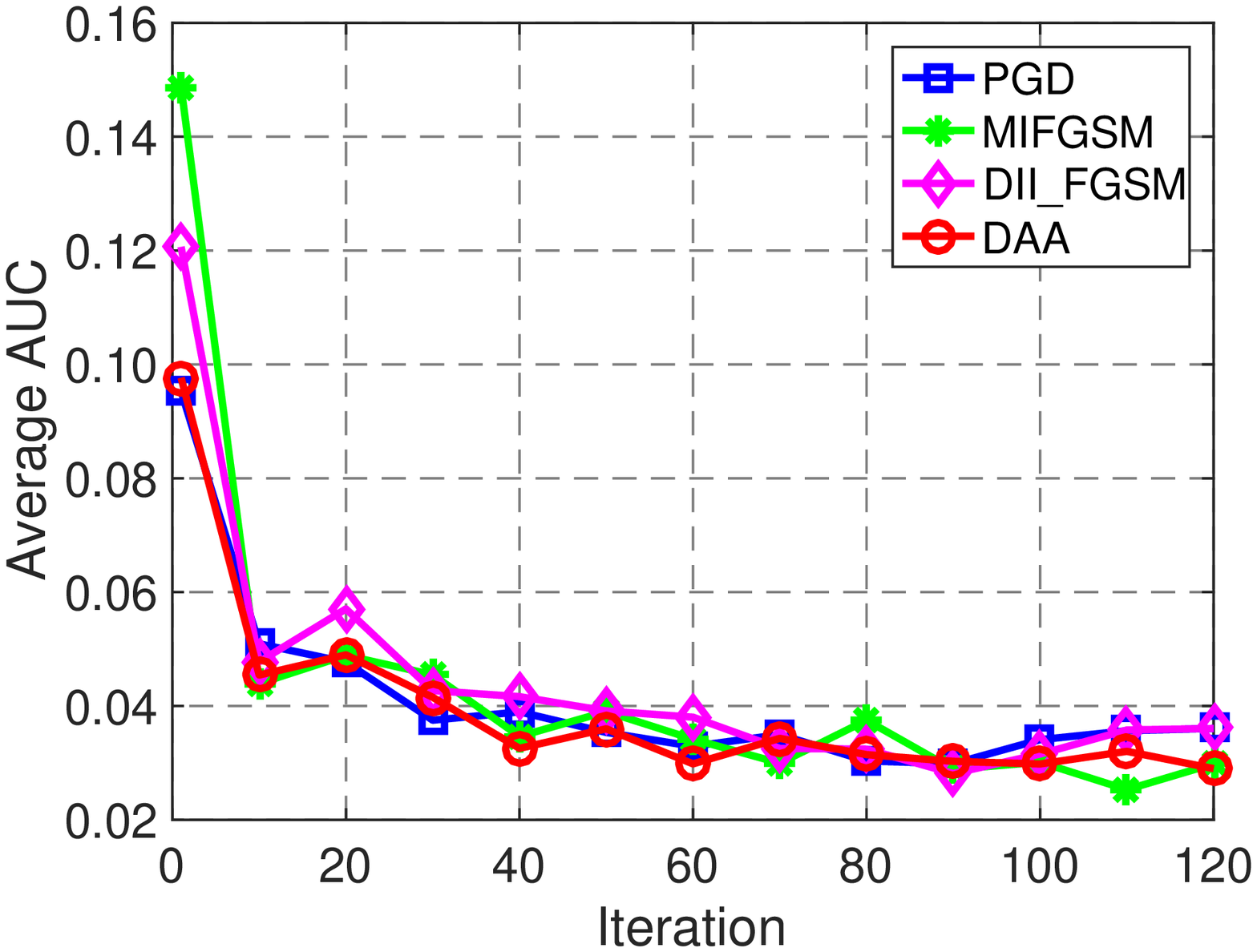}\\
    \end{minipage}%
}%
\subfigure[]{
    \begin{minipage}[t]{0.5\linewidth}
        \centering
        \includegraphics[width=1\linewidth]{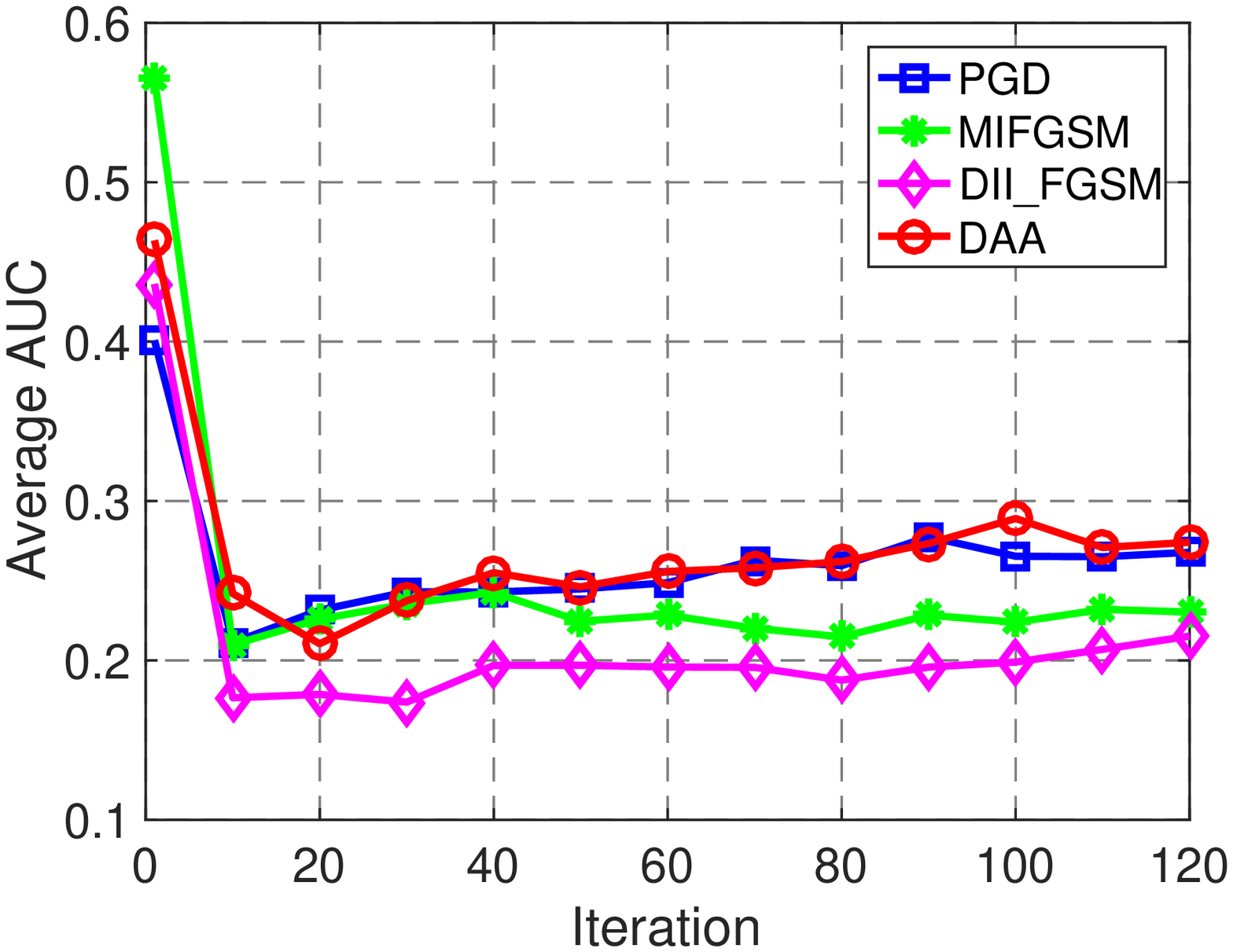}\\
    \end{minipage}%
}
\centering
\caption{(a) shows an average AUC of models for white box attacks that generate adversarial examples using Densenet121. (b) refers to black box attacks that generate adversarial examples using Densenet121 and attack Resnet50 by different numbers of iterations.}

\label{fig:fig4}
\end{figure}

\begin{table}[t]
    \scriptsize
	\centering
    \resizebox{1.0\linewidth}{!}{
	\begin{tabular}{c|c|c|c|c|c}
        \hline
        Method & Setting & Den-121 & Res-50 & VGG16 & Inc-v3 \\
        \hline
		\hline
		\multirow{2}{*}{PGD} & Ensemble&\textbf{0.07746 } & \textbf{0.07956} &\textbf{0.06817} &\textbf{0.0727} \\
		~ & Hold-out&0.10428& 0.14015& 0.35561 &0.35126  \\
		\hline

		\hline
		\multirow{2}{*}{MIFGSM} & Ensemble&\textbf{0.07101 } & \textbf{0.07283} &\textbf{0.05969} &\textbf{0.06690} \\
		~ & Hold-out&0.07325  & 0.13158& 0.33455& 0.33658  \\
		\hline

        \hline
		\multirow{2}{*}{DAA} & Ensemble&\textbf{0.03486} & \textbf{0.04054} &\textbf{0.02825} &\textbf{0.02896} \\
		~ & Hold-out&0.09565 &0.14172 & 0.30138 &0.23225  \\
		\hline

        \hline
		\multirow{2}{*}{DII-FGSM} & Ensemble&\textbf{0.06494} & \textbf{0.08326} &\textbf{0.07077} &\textbf{0.07547} \\
		~ & Hold-out&0.0676&0.12341 &0.14919  &0.20604  \\
		\hline

        \hline
		\multirow{2}{*}{FGSM} & Ensemble&\textbf{0.4909} & \textbf{0.47384} &\textbf{0.55429} &\textbf{0.53963} \\
		~ & Hold-out&0.36769 &0.44520& 0.50624&0.55651 \\
		\hline
	\end{tabular}}
 \caption{ The average AUC of the untargeted adversarial attacks of the ensemble method. We study four models Den121, Res50, VGG16, and Inc-v3. The models in the first row represent the hold-out model. The adversarial examples are generated by the other three models and tested on the ensemble models ($\textbf{white~box~attacks}$) and the hold-out model ($\textbf{black ~box ~attacks}$).}
 \label{tabel:tabel4}
\end{table}

\paragraph{Comparisons of  state-of-the-art attack methods.}

We conduct experiments on both single model and ensemble model using the white box attacks and black box attacks. 
For single models, from Table \ref{tabel:tabel3}, FGSM  has higher AUC than other methods in the white box attacks. DAA outperforms PGD, regardless of whether in white box attacks or black box attacks. In black box attacks, DII-FGSM and MIFGSM outperform PGD. In white box attacks, they also keep high success rates as PGD.  

For ensemble models, from Table \ref{tabel:tabel4}, we generate adversarial examples using the ensemble models and attack the ensemble models and hold-out models in the white box attacks and black box attacks. DAA generates adversarial examples with stronger attack abilities than those of PGD in  both white box attacks and black box attacks. DII-FGSM and MIFGSM  get lower AUC than those of PGD in the black box attacks. In addition, the adversarial examples generated by DII-FGSM with ensemble models have better transferability than that of MIFGSM in the black box attacks.


\paragraph{Impact of iterations number.}
 To study the impact of the number of iterations for attack methods, we conduct experiments with different numbers of iterations in the white box attacks and black box attacks. From Figure \ref{fig:fig4}, when the number of  iterations increases from 0 to 10,  the average AUC  drops sharply. 
The results demonstrate that increasing too much noise in images might not always work well when attacking models. In the white box attacks, the average AUC keeps fluctuating and does not decrease as the number of iterations increases. In the black box attacks, adversarial examples attack other models due to the transferability. The results show that too much noise weakens the transferability of adversarial examples and reduces the success rates of black box attacks.

\begin{figure}[t]

    \subfigure[]{
        \includegraphics[width=0.5\linewidth]{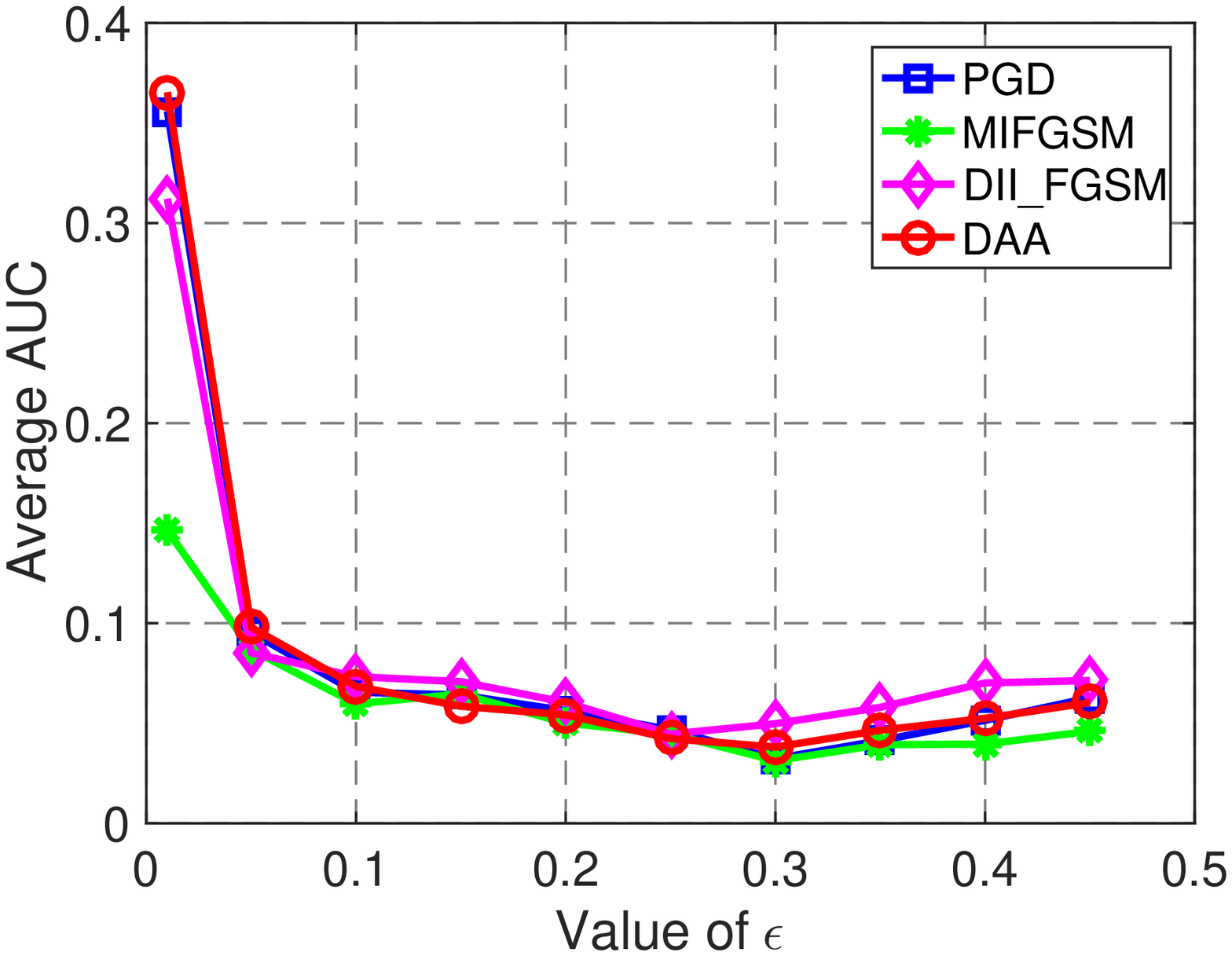}
    }%
    \subfigure[]{
        \includegraphics[width=0.5\linewidth]{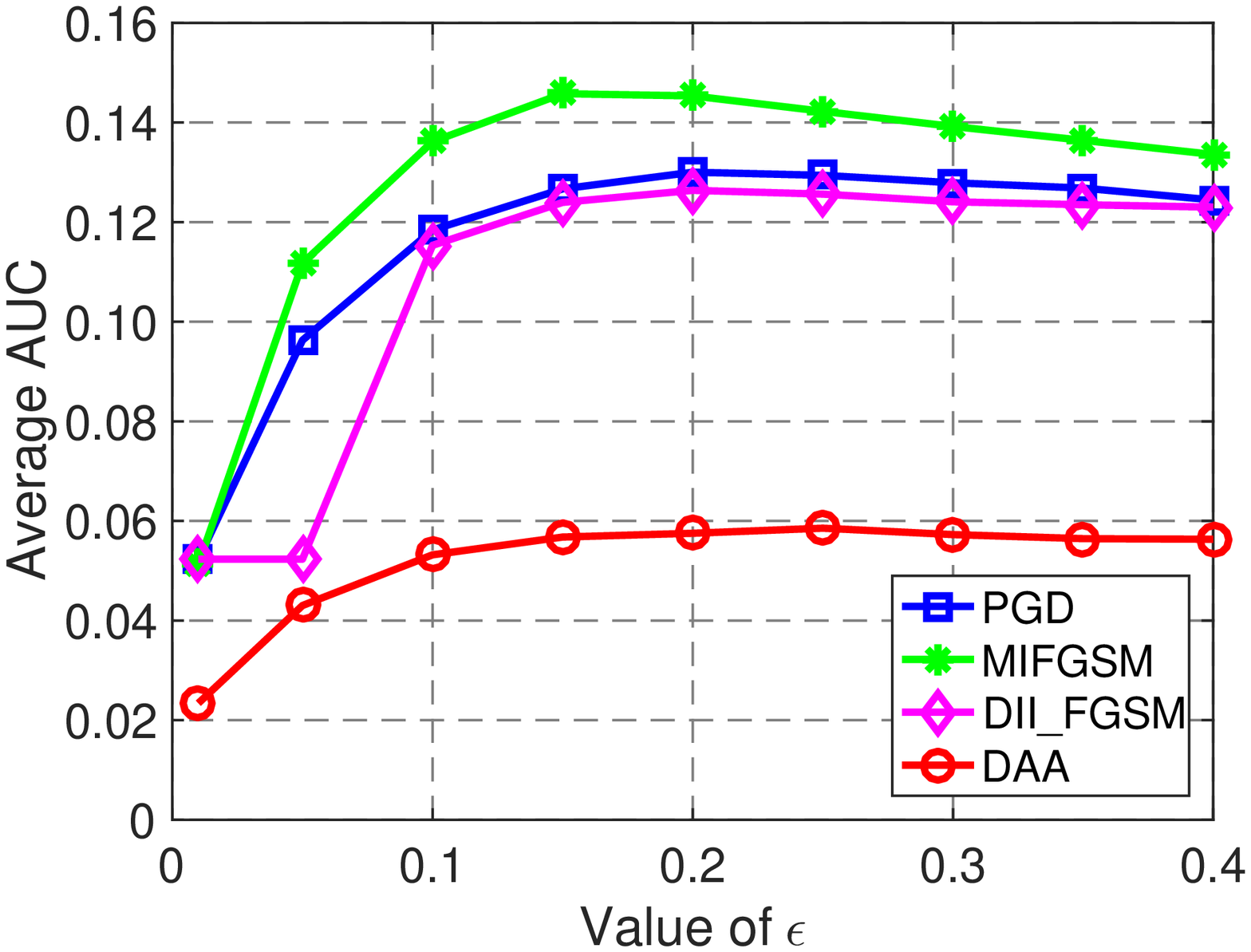}
    }
\centering
\caption{(a) shows average AUC of attacking models with different  $\epsilon$ in the white box attacks, (b) represents the average distance between clean images and adversarial examples. The adversarial examples are generated by Densenet121.  }
\label{fig:fig 5}
\end{figure}

\paragraph{Influence of perturbations size.}
To investigate how the size of perturbations affects adversarial examples, we set the number of iterations as 40 and apply the $l_{2}$ distance to evaluate the distance between the adversarial examples and clean images with different  $\epsilon$. From  Figure \ref{fig:fig 5} (a), the average AUC quickly drops when  $\epsilon$ is smaller than 0.1. When   $\epsilon$ is more than 0.25, the average AUC of the DII-FGSM begins to rise, and the average AUC of PGD and DAA begin to rise when $\epsilon$ is equal to 0.3. From  Figure \ref{fig:fig 5} (b), the distance between adversarial examples and the corresponding clean images steadily rises as $\epsilon$ increases. When $\epsilon$ is large enough, the distance stops growing and even declines.
It also shows that MIFGSM gets the largest distance in all methods.
Thus, too large-sized perturbations do not always lead to good attack performance. When $\epsilon$ is smaller, the distance rises smoothly. When $\epsilon$ increase,  the distance drops and the performance of attack methods decrease.

\subsection{Defense on Multi-label Classification Task}
PGD adversarial training is a method that improves the robustness of models by training with adversarial examples generated by the PGD attack method.
In the experiments, we train Densenet121 using both clean images and adversarial examples. The pixel deflection transform locally corrupts the images by redistributing pixel values. We craft adversarial examples using Densenet121 and input them into Densenet121, Resnet50, VGG16, and Inception v3, respectively.



\paragraph{Results of PGD adversarial training.}

\begin{table}[t] 
	\centering
     \resizebox{0.85\linewidth}{!}{
	\begin{tabular}{c|c|c|c}
		\hline
		  Method    &     Den-PGD$_{multi}$      & Den-121 & Res-50  \\ \hline\hline
		    PGD     & \textbf{0.80735} & 0.85294 & 0.84494 \\ \hline
		  MIFGSM    & \textbf{0.80376} & 0.81709 & 0.80437 \\ \hline
		    DAA     & \textbf{0.83554} & 0.86454& 0.85925\\ \hline
		 DII-FGSM   & \textbf{0.80813} & 0.84649 & 0.8364 \\ \hline
		   FGSM     & \textbf{0.85801} & 0.85185 &0.84008 \\ \hline
	\end{tabular}}
\caption{ The average AUC of untargeted attacks for PGD adversarial training on multi-label classifiers. Den-PGD$_{multi}$ represents the model trained by the PGD adversarial training with Densenet121 as the multi-label classifier. Den-121 represents the Densenet121 trained just using clean images. During the experiments, we set Den-PGD$_{multi}$ as the target models and generate adversarial examples using Den-PGD$_{multi}$, Den-121, and Res-50. 
}
\label{tabel:multi-label}
\end{table}

To evaluate PGD adversarial training, we conduct the experiment on the multi-label classifier using the data set that contains six observations from CheXpert. From the experiment, the average  AUC of clean images is 0.87633, which  still retains a high value. 
Table \ref{tabel:multi-label}  shows the results of adversarial training on the multi-label classifier. 
In both black box attacks and white box attacks, the AUC of all methods is over 0.8. The MIFGSM has the lowest AUC in multi-step attacks, and DAA has the highest AUC in multi-step attacks in both black box attacks and white box attacks.


\begin{table}[t]
	\centering

     \setlength{\tabcolsep}{2.5mm}
	\begin{tabular}{c|c|c|c}
        \hline
          Method & Den-121 & Res-50 & VGG16 \\
        \hline
        \hline

         \hline
          PGD& \textbf{0.68788} &0.69659 & 0.67024  \\
        \hline

        \hline
          MIFGSM& \textbf{0.68143} & 0.69564&0.66539  \\
        \hline

        \hline
          DAA& \textbf{0.68847} & 0.69761&0.67078 \\
        \hline

        \hline
        DII-FGSM& \textbf{0.68538} &0.69683 & 0.66911 \\
        \hline

        \hline
         FGSM& \textbf{0.69355} & 0.70093&0.67431\\
        \hline
	\end{tabular}
 \caption{ The average AUC of the Pixel Deflection Transform. The adversarial examples are generated by Den-121. Den-121, Res-50 and VGG16 are target models.}
  \label{tabel:PDT}
\end{table}

\paragraph{Results of pixel deflecting transform.}
To evaluate the pixel deflecting transform, we conduct an experiment on the multi-label classifier with six observations from CheXpert.
From the experiments, The AUC of the clean images for Densenet121 is 0.70302, which is lower than that of PGD adversarial training.
Table \ref{tabel:PDT} shows the pixel deflection transform results.
 In the white box attacks, AUC is over 0.68. In the black box attacks, AUC also retains high values as in the white box attacks.
It shows that PDT is robust to adversarial examples in both white box attacks and black box attacks, but PDT is less accurate for clean images than PGD adversarial training. In this experiment, PDT just deflects pixels of images without any training procedure, so it consumes less time than adversarial training.

\paragraph{Results of our defense method.}
\begin{figure}[t]
\centering
\includegraphics[width =.95\linewidth]{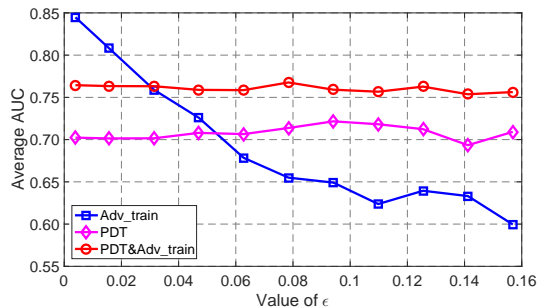}
\caption{The results of different $\epsilon$ for the three defense methods. For PDT and PDT$\&$Adv$\_$train, we generate adversarial examples with PGD using Densenet121. For Adv$\_$train, we generate adversarial examples with PGD using Den-PGD$_{multi}$.}
\label{fig:fig6}
\end{figure}
In this part, we combine the pixel deflecting transform with PGD adversarial training (PDT$\&$Adv$\_$train). First, we train Densenet121 using the multi-label classifier with samples that combine adversarial examples and clean images. Then, we process the images using PDT and input them into the network trained by PGD adversarial training. At last, we compare the methods with PDT and PGD adversarial training, respectively, with different sized perturbations. We show the results in Figure \ref{fig:fig6}.

Figure~\ref{fig:fig6} shows PDT$\&$Adv$\_$train is superior to PDT with various $\epsilon$. As $\epsilon$ increases, AUC of Adv$\_$train declines and AUC of PDT and PDT$\&$Adv$\_$train remain steady. Adv$\_$train has the highest AUC over other two methods when $\epsilon$ is smaller than 0.03. PDT$\&$Adv$\_$train outperforms Adv$\_$train when $\epsilon$ is over 0.03.
The results demonstrate that PGD adversarial training and PDT are able to defend against adversarial examples. PGD adversarial training retains more accurate clean images than those of PDT. Moreover, when $\epsilon$ increases, AUC of PGD adversarial training drops sharply. Combining PGD adversarial training and PDT achieves better performance than that of PDT and keeps AUC steady as $\epsilon$ increases.


\section{Discussions and Challenges}

In this section, we will discuss the challenges for attacks and defense when analyzing chest X-rays. Moreover, we discuss the perturbations and robustness of models.

\paragraph{Attack and defense on multi-label classification.}
Most attack and defense methods achieve good performance on multi-class classification tasks.
However, multi-label classification is more difficult than multi-class classification.
Therefore, it is more challenging for attack and defense methods in multi-label classification tasks. 
In the experiment, we apply the targeted attack methods  \cite{song2018multi,xie2018improving,dong2018boosting} on medical images on the multi-label classification task.
However, these methods fail to achieve good performance. 
The reason lies in that these methods are not strong enough in targeted attacks or these methods tested on multi-class classifiers are not suitable for multi-label classifiers in targeted attacks.

 \paragraph{Perturbations and robustness.}
 In Figure \ref{fig:fig6}, when the size of perturbations increases, the average AUC  drops rapidly in PGD adversarial training. \cite{salimans2017pixelcnn++} showed that the distributions of clean images are different from adversarial examples. 
 In this sense, small perturbations added into the image can change the distribution of the image. Therefore, when we increase the size of the perturbations, the distributions of the adversarial examples may be different with samples  used to conduct adversarial training. 
 Furthermore, models trained with adversarial examples with a certain sized perturbations may not be robust to other adversarial examples generated by other sized perturbations.

 For chest X-rays, most attack and defense methods rarely focus on the influence of different degrees of perturbations.
 In practice, we found that the degrees of perturbations affects the robustness of the classifier.
 Therefore, how to improve the robustness of the classifier to defense against different degrees of perturbation remains an open question.
 
\section{Conclusion}
In this paper, we reviewed various adversarial attack and defense methods on chest X-rays and compared their motivations and mathematical representations.
Moreover, we performed a thorough comparison study on the state-of-the-art attack and defense methods for common thorax disease classification in chest X-rays.
This review revealed a limitation of the existing attack and defense methods that they have poor performance with excessive iterations and large perturbation.
To address this limitation, we proposed a robust defense method and evaluated its effectiveness through extensive experiments.
More critically, this review could open a research direction for developing more effective methods to defense against large perturbation and improve the robustness of the models.




\bibliography{example_paper}

\begin{thebibliography}{37}
\providecommand{\natexlab}[1]{#1}
\providecommand{\url}[1]{\texttt{#1}}
\expandafter\ifx\csname urlstyle\endcsname\relax
  \providecommand{\doi}[1]{doi: #1}\else
  \providecommand{\doi}{doi: \begingroup \urlstyle{rm}\Url}\fi

\bibitem[Brown et~al.(2017)Brown, Man{\'e}, Roy, Abadi, and
  Gilmer]{brown2017adversarial}
Brown, T.~B., Man{\'e}, D., Roy, A., Abadi, M., and Gilmer, J.
\newblock Adversarial patch.
\newblock \emph{arXiv}, 2017.

\bibitem[Cernazanu-Glavan \& Holban(2013)Cernazanu-Glavan and
  Holban]{cernazanu2013segmentation}
Cernazanu-Glavan, C. and Holban, S.
\newblock Segmentation of bone structure in x-ray images using convolutional
  neural network.
\newblock \emph{Adv. Electr. Comput. Eng}, 2013.

\bibitem[Dong et~al.(2018)Dong, Liao, Pang, Su, Zhu, Hu, and
  Li]{dong2018boosting}
Dong, Y., Liao, F., Pang, T., Su, H., Zhu, J., Hu, X., and Li, J.
\newblock Boosting adversarial attacks with momentum.
\newblock In \emph{CVPR}, 2018.

\bibitem[Goodfellow et~al.(2014)Goodfellow, Shlens, and
  Szegedy]{goodfellow2014explaining}
Goodfellow, I.~J., Shlens, J., and Szegedy, C.
\newblock Explaining and harnessing adversarial examples.
\newblock \emph{arXiv}, 2014.

\bibitem[Guo et~al.(2017)Guo, Rana, Cisse, and Van
  Der~Maaten]{guo2017countering}
Guo, C., Rana, M., Cisse, M., and Van Der~Maaten, L.
\newblock Countering adversarial images using input transformations.
\newblock \emph{arXiv}, 2017.

\bibitem[Guo et~al.(2019)Guo, Zheng, Tan, Chen, Chen, Zhao, and
  Huang]{guo2019nat}
Guo, Y., Zheng, Y., Tan, M., Chen, Q., Chen, J., Zhao, P., and Huang, J.
\newblock Nat: Neural architecture transformer for accurate and compact
  architectures.
\newblock In \emph{Advances in Neural Information Processing Systems}, pp.\
  735--747, 2019.

\bibitem[He et~al.(2016)He, Zhang, Ren, and Sun]{he2016deep}
He, K., Zhang, X., Ren, S., and Sun, J.
\newblock Deep residual learning for image recognition.
\newblock In \emph{CVPR}, 2016.

\bibitem[Hinton et~al.(2015)Hinton, Vinyals, and Dean]{hinton2015distilling}
Hinton, G., Vinyals, O., and Dean, J.
\newblock Distilling the knowledge in a neural network.
\newblock \emph{arXiv}, 2015.

\bibitem[Hu et~al.(2019)Hu, Yu, Guo, Chao, and Weinberger]{hu2019new}
Hu, S., Yu, T., Guo, C., Chao, W.-L., and Weinberger, K.~Q.
\newblock A new defense against adversarial images: Turning a weakness into a
  strength.
\newblock In \emph{Advances in Neural Information Processing Systems}, pp.\
  1633--1644, 2019.

\bibitem[Huang et~al.(2017)Huang, Liu, Van Der~Maaten, and
  Weinberger]{huang2017densely}
Huang, G., Liu, Z., Van Der~Maaten, L., and Weinberger, K.~Q.
\newblock Densely connected convolutional networks.
\newblock In \emph{CVPR}, 2017.

\bibitem[Irvin et~al.(2019)Irvin, Rajpurkar, Ko, Yu, Ciurea-Ilcus, Chute,
  Marklund, Haghgoo, Ball, Shpanskaya, et~al.]{irvin2019chexpert}
Irvin, J., Rajpurkar, P., Ko, M., Yu, Y., Ciurea-Ilcus, S., Chute, C.,
  Marklund, H., Haghgoo, B., Ball, R., Shpanskaya, K., et~al.
\newblock Chexpert: A large chest radiograph dataset with uncertainty labels
  and expert comparison.
\newblock In \emph{AAAI}, 2019.

\bibitem[Kurakin et~al.(2016{\natexlab{a}})Kurakin, Goodfellow, and
  Bengio]{kurakin2016adversarial}
Kurakin, A., Goodfellow, I., and Bengio, S.
\newblock Adversarial examples in the physical world.
\newblock \emph{arXiv}, 2016{\natexlab{a}}.

\bibitem[Kurakin et~al.(2016{\natexlab{b}})Kurakin, Goodfellow, and
  Bengio]{kurakin2016adversarial2}
Kurakin, A., Goodfellow, I., and Bengio, S.
\newblock Adversarial machine learning at scale.
\newblock \emph{arXiv}, 2016{\natexlab{b}}.

\bibitem[Lakhani \& Sundaram(2017)Lakhani and Sundaram]{lakhani2017deep}
Lakhani, P. and Sundaram, B.
\newblock Deep learning at chest radiography: automated classification of
  pulmonary tuberculosis by using convolutional neural networks.
\newblock \emph{Radiology}, 2017.

\bibitem[Li et~al.(2019)Li, Yi, Zhou, and Zhang]{li2019improving}
Li, P., Yi, J., Zhou, B., and Zhang, L.
\newblock Improving the robustness of deep neural networks via adversarial
  training with triplet loss.
\newblock \emph{arXiv}, 2019.

\bibitem[Madry et~al.(2018)Madry, Makelov, Schmidt, Tsipras, and
  Vladu]{madry2017towards}
Madry, A., Makelov, A., Schmidt, L., Tsipras, D., and Vladu, A.
\newblock Towards deep learning models resistant to adversarial attacks.
\newblock \emph{ICLR}, 2018.

\bibitem[Moosavi-Dezfooli et~al.(2016)Moosavi-Dezfooli, Fawzi, and
  Frossard]{moosavi2016deepfool}
Moosavi-Dezfooli, S.-M., Fawzi, A., and Frossard, P.
\newblock Deepfool: a simple and accurate method to fool deep neural networks.
\newblock In \emph{CVPR}, 2016.

\bibitem[Pasa et~al.(2019)Pasa, Golkov, Pfeiffer, Cremers, and
  Pfeiffer]{pasa2019efficient}
Pasa, F., Golkov, V., Pfeiffer, F., Cremers, D., and Pfeiffer, D.
\newblock Efficient deep network architectures for fast chest x-ray
  tuberculosis screening and visualization.
\newblock \emph{Scientific reports}, 2019.

\bibitem[Prakash et~al.(2018)Prakash, Moran, Garber, DiLillo, and
  Storer]{prakash2018deflecting}
Prakash, A., Moran, N., Garber, S., DiLillo, A., and Storer, J.
\newblock Deflecting adversarial attacks with pixel deflection.
\newblock In \emph{CVPR}, 2018.

\bibitem[Rajpurkar et~al.(2017)Rajpurkar, Irvin, Zhu, Yang, Mehta, Duan, Ding,
  Bagul, Langlotz, Shpanskaya, et~al.]{rajpurkar2017chexnet}
Rajpurkar, P., Irvin, J., Zhu, K., Yang, B., Mehta, H., Duan, T., Ding, D.,
  Bagul, A., Langlotz, C., Shpanskaya, K., et~al.
\newblock Chexnet: Radiologist-level pneumonia detection on chest x-rays with
  deep learning.
\newblock \emph{arXiv}, 2017.

\bibitem[Salimans et~al.(2017)Salimans, Karpathy, Chen, and
  Kingma]{salimans2017pixelcnn++}
Salimans, T., Karpathy, A., Chen, X., and Kingma, D.~P.
\newblock Pixelcnn++: Improving the pixelcnn with discretized logistic mixture
  likelihood and other modifications.
\newblock \emph{arXiv}, 2017.

\bibitem[Samangouei et~al.(2018)Samangouei, Kabkab, and
  Chellappa]{samangouei2018defense}
Samangouei, P., Kabkab, M., and Chellappa, R.
\newblock Defense-gan: Protecting classifiers against adversarial attacks using
  generative models.
\newblock \emph{arXiv}, 2018.

\bibitem[Saul et~al.(2019)Saul, Urey, and Taktakoglu]{saul2019early}
Saul, C.~J., Urey, D.~Y., and Taktakoglu, C.~D.
\newblock Early diagnosis of pneumonia with deep learning.
\newblock \emph{arXiv}, 2019.

\bibitem[Shi et~al.(2019)Shi, Wang, and Han]{shi2019curls}
Shi, Y., Wang, S., and Han, Y.
\newblock Curls \& whey: Boosting black-box adversarial attacks.
\newblock In \emph{Proceedings of the IEEE Conference on Computer Vision and
  Pattern Recognition}, pp.\  6519--6527, 2019.

\bibitem[Simonyan \& Zisserman(2014)Simonyan and Zisserman]{simonyan2014very}
Simonyan, K. and Zisserman, A.
\newblock Very deep convolutional networks for large-scale image recognition.
\newblock \emph{arXiv}, 2014.

\bibitem[Song et~al.(2018)Song, Jin, Huang, and Hu]{song2018multi}
Song, Q., Jin, H., Huang, X., and Hu, X.
\newblock Multi-label adversarial perturbations.
\newblock In \emph{ICDM}, 2018.

\bibitem[Stephen et~al.(2019)Stephen, Sain, Maduh, and
  Jeong]{stephen2019efficient}
Stephen, O., Sain, M., Maduh, U.~J., and Jeong, D.-U.
\newblock An efficient deep learning approach to pneumonia classification in
  healthcare.
\newblock \emph{Journal of healthcare engineering}, 2019.

\bibitem[Su et~al.(2019)Su, Vargas, and Sakurai]{su2019one}
Su, J., Vargas, D.~V., and Sakurai, K.
\newblock One pixel attack for fooling deep neural networks.
\newblock \emph{IEEE Transactions on Evolutionary Computation}, 2019.

\bibitem[Szegedy et~al.(2013)Szegedy, Zaremba, Sutskever, Bruna, Erhan,
  Goodfellow, and Fergus]{szegedy2013intriguing}
Szegedy, C., Zaremba, W., Sutskever, I., Bruna, J., Erhan, D., Goodfellow, I.,
  and Fergus, R.
\newblock Intriguing properties of neural networks.
\newblock \emph{arXiv}, 2013.

\bibitem[Szegedy et~al.(2016)Szegedy, Vanhoucke, Ioffe, Shlens, and
  Wojna]{szegedy2016rethinking}
Szegedy, C., Vanhoucke, V., Ioffe, S., Shlens, J., and Wojna, Z.
\newblock Rethinking the inception architecture for computer vision.
\newblock In \emph{CVPR}, 2016.

\bibitem[Tram{\`e}r et~al.(2017)Tram{\`e}r, Kurakin, Papernot, Goodfellow,
  Boneh, and McDaniel]{tramer2017ensemble}
Tram{\`e}r, F., Kurakin, A., Papernot, N., Goodfellow, I., Boneh, D., and
  McDaniel, P.
\newblock Ensemble adversarial training: Attacks and defenses.
\newblock \emph{arXiv}, 2017.

\bibitem[Wang \& Xia(2018)Wang and Xia]{wang2018chestnet}
Wang, H. and Xia, Y.
\newblock Chestnet: A deep neural network for classification of thoracic
  diseases on chest radiography.
\newblock \emph{arXiv}, 2018.

\bibitem[Wu et~al.(2019)Wu, Phang, Park, Shen, Huang, Zorin, Jastrzebski,
  Fevry, Katsnelson, Kim, et~al.]{wu2019deep}
Wu, N., Phang, J., Park, J., Shen, Y., Huang, Z., Zorin, M., Jastrzebski, S.,
  Fevry, T., Katsnelson, J., Kim, E., et~al.
\newblock Deep neural networks improve radiologists’ performance in breast
  cancer screening.
\newblock \emph{IEEE Transactions on Medical Imaging}, pp.\  1--1, 2019.

\bibitem[Xie et~al.(2018)Xie, Zhang, Wang, Zhou, Ren, and
  Yuille]{xie2018improving}
Xie, C., Zhang, Z., Wang, J., Zhou, Y., Ren, Z., and Yuille, A.
\newblock Improving transferability of adversarial examples with input
  diversity.
\newblock \emph{arXiv}, 2018.

\bibitem[Zhang et~al.(2019{\natexlab{a}})Zhang, Chen, Wei, Zhao, Cao,
  et~al.]{zhang2019whole}
Zhang, Y., Chen, H., Wei, Y., Zhao, P., Cao, J., et~al.
\newblock From whole slide imaging to microscopy: Deep microscopy adaptation
  network for histopathology cancer image classification.
\newblock In \emph{International Conference on Medical Image Computing and
  Computer-Assisted Intervention}, pp.\  360--368. Springer,
  2019{\natexlab{a}}.

\bibitem[Zhang et~al.(2019{\natexlab{b}})Zhang, Wei, Zhao, Niu, Wu, Tan, and
  Huang]{zhang2019collaborative}
Zhang, Y., Wei, Y., Zhao, P., Niu, S., Wu, Q., Tan, M., and Huang, J.
\newblock Collaborative unsupervised domain adaptation for medical image
  diagnosis.
\newblock In \emph{Medical Imaging meets NeurIPS}, 2019{\natexlab{b}}.

\bibitem[Zheng et~al.(2018)Zheng, Chen, and Ren]{zheng2018distributionally}
Zheng, T., Chen, C., and Ren, K.
\newblock Distributionally adversarial attack.
\newblock \emph{arXiv}, 2018.

\end{thebibliography}
\bibliographystyle{icml2020}

\end{document}